\documentclass[12pt]{article}

\usepackage{color}
\definecolor{lightblue}{rgb}{0,0.2,0.5}

\usepackage[utf8]{inputenc} 
\usepackage[T1]{fontenc}    

\usepackage[colorlinks=true, urlcolor=lightblue,linkcolor=lightblue, citecolor=lightblue]{hyperref}
\usepackage{url}            
\usepackage{booktabs}       
\usepackage{amsfonts}       
\usepackage{nicefrac}       
\usepackage{microtype}      
\usepackage{amsmath}
\usepackage{relsize}
\usepackage{xcolor}
\usepackage{soul}
\usepackage{algorithm}
\usepackage{algpseudocode}
\usepackage{graphicx}
\graphicspath{ {./images/} }
\usepackage{listings} 
\usepackage{multirow}
\usepackage{makecell}

\usepackage{enumerate}

\usepackage{makecell}
\usepackage{threeparttable}
\usepackage[german=guillemets]{csquotes}

\usepackage{dsfont} 

\usepackage{doi} 

\newtheorem{prop}{Proposition}[section]

\newtheorem{lemma}[prop]{Lemma}

\newenvironment{Proof}{\removelastskip\par\medskip
\noindent{\em Proof.} \rm}{\penalty-20\null\hfill$\square$\par\medbreak}

\usepackage[round]{natbib}

\def\real{{\mathord{\mathbb R}}}

\title{\huge Deep self-consistent learning of local volatility}

\author{Zhe~Wang \hskip1cm Ameir Shaa\footnote{Starting 9 Dec 2024, the second author will be with School of Physical and Mathematical Sciences, Nanyang Technological University, 637371 Singapore.}
  \\
  \small
  CNRS@CREATE Ltd
  \\
  \small
 \#08-01 CREATE Tower
  \\
  \small
 1 Create Way, Singapore 138602
  \bigskip
  \\
 Nicolas~Privault \hskip1cm Claude Guet
  \\
  \small
  School of Physical and Mathematical Sciences
  \\
  \small
  Nanyang Technological University
  \\
  \small
  21 Nanyang Link, 637371 Singapore.
  \medskip
}

\begin{document}

\maketitle

\vspace{-1cm}

\begin{abstract}
We present an algorithm for the calibration of local volatility from market option prices through deep self-consistent learning, by approximating both market option prices and local volatility using deep neural networks. Our method uses the initial-boundary value problem of the underlying Dupire's partial differential equation solved by the parameterized option prices to bring corrections to the parameterization in a self-consistent way. By exploiting the differentiability of neural networks, we can evaluate Dupire's equation locally at each strike-maturity pair; while by exploiting their continuity, we sample strike-maturity pairs uniformly from a given domain, going beyond the discrete points where the options are quoted. Moreover, the absence of arbitrage opportunities are imposed by penalizing an associated loss function as a soft constraint. For comparison with existing approaches, the proposed method is tested on both synthetic and market option prices, which shows an improved performance in terms of reduced interpolation and reprice errors, as well as the smoothness of the calibrated local volatility. An ablation study has been performed, asserting the robustness and significance of the proposed method.
\end{abstract}


\baselineskip0.7cm

\newpage

\section{Introduction}

 An option is a financial contract that gives options holders the right but not the obligation to buy (a call option) or to sell (a put option) an asset, e.g. stocks or commodity, for a predetermined price (strike price) on a predetermined date (maturity). To acquire today the right to buy or sell an asset in the future, a premium, namely the option price, must be paid to the option writers. The standard approach to model option prices is based on the Black-Scholes formula, which assumes a constant local volatility surface and therefore applies only to situations where the stock prices process follows a geometric Brownian motion.
However, the market price data
of underlying assets typically invalidate the assumptions of
standard Brownian motion, hence a more sophisticated model than the Black-Scholes formula is required. Among alternatives, we consider an asset price process expressed in a local volatility model as 
\begin{equation} \label{eqn BS} 
 dS_t = r S_t dt + \sigma(S_t, t) S_t dB_t, \quad {S_t}_{\vert t=0} = S_0.
\end{equation} 
where $S_t$ denotes the stock price at the time $t \in \mathbb{R}_{+}$,
and $(B_t)_{t\in \real_+}$ is a standard Brownian motion. For simplicity, we assume a zero dividend yield and a constant risk-free interest rate $r$.
Rather than taking a constant value as in the canonical Black-Scholes equation, in Eq.~\eqref{eqn BS}, the local volatility $\sigma(S_t, t)$ is a deterministic function of the underlying asset price $S_t$ and of time $t$, satisfying the usual Lipschitz conditions. Without introducing additional sources of randomness, the local volatility model is the only complete and consistent model that allows hedging based on solely the underlying asset, 
cf. Appendix A1 in \cite{Bennett14}; and it is used  for daily risk management
 in most 
 investment bank production systems. 

The practical implementation of such a stochastic volatility model in option pricing requires to solve the challenging problem of calibrating the local volatility function to the market data of option prices. Given $\pi(K,T)$ a family of option prices with strike prices $K>0$, maturities $T>0$, and underlying asset price $S_0=x$, the Dupire formula \cite{dupire,derman-kani} 
\begin{equation} 
\label{eqn Dupire_formula}
 \sigma (K, T) : = \sqrt{ 
 \displaystyle 
 \frac{ 
 \displaystyle 
 2 \frac{\partial \pi}{\partial T} (K, T) 
 + 
 2 r K \frac{\partial \pi}{\partial K} (K, T) 
 }{ 
 \displaystyle 
 K^2 
 \frac{\partial^2 \pi}{\partial K^2} (K, T) 
 }}
, 
\end{equation}
brings a solution to this problem by constructing an estimator of the local volatility as a function of strike price $K$ and maturity $T$ values, i.e. Eq.~\eqref{eqn Dupire_formula} matches the stock's volatility $\sigma (S_t, t)$ when the
 underlying stock price $S_t$ is at the level $K$ at time $T=t$. 

The numerical estimation of local volatility using Eq.~\eqref{eqn Dupire_formula}, involves evaluating partial derivatives, which is classically achieved using the finite difference method. More efficient methods have been introduced using spline functions, see e.g. Chapter~8 in \cite{Achdou05}, where local volatility is first approximated as the sum a piecewise affine function satisfying a boundary condition and a bi-cubic spline function. Then, the model parameters are determined by minimizing the discretized Dupire's equations \cite{dupire} at selected collocation points via Tikhonov regularization. Thereafter, Tikhonov regularization has been applied to the calibration of local volatility in a trinomial model in \cite{Crepey02}. 

Alternative to calibrating local volatility using parameterized functions, neural networks are known being able to approximate any arbitrary nonlinear function with a finite set of parameters \cite{Gorban98,Winkler17,Lin18}. By exploiting this property, Chataigner et al. \cite{Chataigner20} advocated to parameterize the market option prices $\pi(K, T)$ using neural networks, which allow for the computation of the derivatives in Eq.~\eqref{eqn Dupire_formula} by automatic differentiation \cite{Baydin17}.

In addition, for Dupire's formula to be meaningful, one must ensure that the argument in the square root is positive and remains bounded in \eqref{eqn Dupire_formula}, which can be achieved under a no arbitrage condition for the quoted option prices.
In \cite{Chataigner20},
additional constraints are imposed in order to
ensure that the market option prices are well fitted,
in which case an approximation of derivatives of the option price is then substituted into Eq.~\eqref{eqn Dupire_formula} to calibrate the local volatility.

In this case, the quality of the reconstructed local volatility surface depends on the resolution and quality of the quoted market option prices at given strike-maturity pairs. Hence, instead of using market option price, Chataigner et al. \cite{Chataigner21} proposed subsequently, to use the implied volatility surface for the calibration of local volatility, leading to an improved performance. It is, however, important to note that, when fitting option prices or implied volatilities using neural networks, one assumes that the data are noise free. Except for those that violate the arbitrage-free conditions, the proposed methods in \cite{Chataigner20,Chataigner21} are not able to filter out noises in the data. 

In this work, we follow the approach of \cite{WangZhe22}, and calibrate the local volatility surface from the observed market option prices in a self-consistent manner. Going beyond merely fitting option prices using neural networks, we rely on recent studies on physics-informed machine learning \cite{Rassi19,Rackauckas20,Karniadakis21,WangZhe22} which have shown that the inclusion of the residue of an unknown, underlying governing equation as a regularizer can not only calibrate the unknown terms in the governing equations, but also bring correction to the data by filtering out noises that are not characterized by the equation.

For this, we approximate both the market option prices and the squared local volatility using deep neural networks. By requiring $\pi(K, T)$ to be a solution to the Dupire equation subjected to given initial and boundary conditions to be discussed in Section~\ref{sec. dupire_eqn}, we regularize $\pi(K, T)$ and determine the unknown $\sigma(K, T)$ self-consistently. Since the resulting $\pi(K, T)$ and $\sigma(K, T)$ are continuous and differentiable functions of $K$ and $T$, we can evaluate Dupire's equation locally on a uniformly sampled $(K, T)$ from the support of Dupire's equation. Unlike previous approaches, where constraints were imposed on fixed strike-maturity pairs, a successive re-sampling enables us to regularize the entire support. In addition, the positiveness of $\sigma(K, T)$ is ensured by a properly selected output activation function of the neural network, hence it is everywhere well defined.  Finally, we mitigate the risk of arbitrage opportunities by penalizing a loss as soft constraints. 
\label{jkldf1}
 In this way, local volatility is estimated by
 construction as a smooth surface instead of applying 
 automatic differentiation to neural option prices.
 In addition, noise in market data is filtered out from the
 constraint imposed by Dupire's equation. 
 A TensorFlow implementation of our algorithm is available
 at 
 \url{https://github.com/ameirtheshaa/LocalVolatility}.
           
 The rest of the paper is organized as follows. In Section~\ref{s2} we review background knowledge on Dupire's equation and arbitrage-free conditions, followed by a comment on market data. After rescaling and reparameterization of market option prices in Section~\ref{sec:change_of_variable} and a discussion of neural ansatz and loss functions in Section~\ref{sec:loss}, our deep self-consistent algorithm is summarized and discussed in Section~\ref{sec:algo}. The proposed method is first tested on synthetic datasets in Section~\ref{sec:synthetic_data}, and then applied to market option prices in Section~\ref{sec:real_data} where, as an ablation study, we compare our results with those obtained without including the residue of Dupire's equation as a regularizer. Finally, the advantages and limitations of our method are discussed in Section~\ref{sec:limitations}, where conclusions are drawn.

\section{Background and objectives} 
\label{s2}

\subsection{Dupire's equation}
\label{sec. dupire_eqn}

Denoting by $\pi^c(K,T)$ and $\pi^p(K,T)$ the prices for European call and put options
\begin{equation} \label{eqn expectation_option}
  \pi^c(K,T) = e^{-rT} \mathbb{E} \big[ ( S_T - K )^+  \big],
  \qquad
    \pi^p(K,T) = e^{-rT} \mathbb{E}\big[ ( K - S_T )^+  \big],
     \end{equation}
the relation between the market option price and the local volatility is established explicitly by Dupire's equation \cite{dupire}
\begin{align}
  \label{eqn dupire_original}
    \frac{\partial\pi}{\partial T}(K, T) + rK \frac{\partial \pi}{\partial K}(K,T) - \frac{1}{2}K^2 \sigma^2(K,T) \frac{\partial^2 \pi}{\partial K^2} (K,T) = 0, 
\end{align}
subject to the initial conditions 
\begin{align}
  \nonumber 
    \pi^c(K,0) = \left( S_0 - K \right)^+, \qquad \pi^p(K,0) = \left( K - S_0 \right)^+ ,
\end{align}
 and to the boundary conditions 
\begin{align}
\nonumber
   \pi^c(\infty, T) = 0, \qquad \quad \pi^p(0, T) = 0, 
\end{align}
respectively for call and put options.
We refer the reader to Chapter 1 in \cite{Itkin20} or Proposition~9.2 in \cite{privaultbkf2} for derivations of Dupire's equation, and to Chapter 8 in \cite{Achdou05} for a discussion of initial and boundary conditions.
From their definitions \eqref{eqn expectation_option}, we note that the prices of European options are nonnegative, with the prices of call options being always less than the value of the underlying asset; while the prices of the put options being always less than the present value of the strike price, leading to the
 lower and upper bounds
 \begin{align}
   \label{eqn inequality}
    0 \leq \pi^c(K, T) \leq S_0, \qquad 0 \leq \pi^p(K,T) \leq K.
\end{align}

\subsection{Arbitrage-free surface}

A static arbitrage is a trading strategy that has zero initial cost and constantly non-negative value afterwards, representing a risk-free profitable investment. 
Under the assumption that economic agents are rational, arbitrage opportunities, if they ever exist, will be instantaneously exploited until the market is arbitrage free. Therefore, as a prerequisite, the absence of arbitrage opportunities is a fundamental principle underpinning the modern theory of financial asset pricing. An option price surface is free of static arbitrage if and only if
\begin{enumerate}[(i)]
\item
  each strike slice is free of calendar spread arbitrage; and
\item
  each maturity slice is free of butterfly arbitrage \cite{Ackerer20,Chataigner20,Chataigner21}.
\end{enumerate}
Next, we consider both calendar spread and
butterfly arbitrages that can be created with put options as well as call options, cf. Sec. 9.2 in \cite{Hull03} for details.
 Calendar spread arbitrages are created by buying a long-maturity put option and selling a short-maturity put option ,
and their absence can be characterized in the continuous limit by the condition
\begin{align}
  \nonumber
  \frac{\partial \pi}{\partial T}(K, T) \geq 0.
\end{align}
On the other hand,
butterfly arbitrages can be created by buying two options at prices $K_1$ and $K_3$, with $K_3 > K_1$; and selling two options with a strike price, $K_2 = (K_1 + K_3)/2$. In the continuous limit, the absence of butterfly arbitrage
is formulated as 
\begin{align}
\nonumber
      \frac{\partial^2 \pi}{\partial K^2} (K,T) \geq 0.
\end{align}

\subsection{Problem formulation}

Market option prices are observed on $N$ discrete pairs of strike price and maturity values, leading to triplets $(\pi_i, K_i, T_i)$ with $i = 1, ..., N$. In practice, to secure a transaction, two option prices are quoted in the market, a bid price and an ask price, and there is no guarantee that the mid-price $\pi_i$ between the bid and ask prices is arbitrage-free \cite{Ackerer20}. Moreover, the observed bid and ask prices may not be updated timely, hence not actionable, leading to additive noise to the market data. Lastly, market option prices are not evenly quoted over a plane spanned by $K$ and $T$: the market data are usually dense close to the money; whereas sparse away from the money. Canonical methods often lack the ability to interpolate option prices and calibrate the associated local volatility without quotes, agilely. Therefore, a validation, a correction, and an interpolation to the market data are required.

Consequently, the main challenge when calibrating the local volatility is twofold. First, given limited observations of unevenly distributed option prices quoted on discrete pairs of strikes and maturities, the calibrated local volatility should span a continuous domain. Second, the corresponding option prices should preclude any arbitrage opportunities. Our objective is, therefore, to construct from observed market option prices a self-consistent approximation for both $\sigma(K,T)$ and $\pi(K,T)$ in the sense that the calibrated local volatility will generate an arbitrage-free option price surface that is in line with the market data up to some physics-informed corrections. 
 
\section{Methodology} 

\subsection{Rescaling Dupire's equation} \label{sec:change_of_variable}
In this section, a change of variable and a rescaling is applied to Dupire's equation \eqref{eqn dupire_original}
to ensure that the terms in the scaled Eq.~\eqref{eqn Dupire_pde_dimensionless} below are
of order of unity, so that no term is dominating over the others during the minimization. For this, we use the change of variables
\begin{align}
  \label{eqn change_of_variables}
    k = e^{-rT} K / K_\text{max}, \quad t = T / T_\text{max},
\end{align}
by picking sufficiently large
 $K_\text{max} = \text{max}(K)$ and $T_\text{max} = \text{max}(T)$. Letting 
\begin{align}
  \nonumber 
 \eta (k,t) := \frac{T_\text{max}}{2} \sigma^2(K, T)
\end{align}
denote the squared local volatility, the Dupire equation (\ref{eqn dupire_original}) can be rewritten as 
\begin{align}
  \label{eqn Dupire_pde_dimensionless}
    f_\text{dup}(k,t) := \frac{\partial {\pi}}{\partial t} (k ,t) - {\eta} (k,t) k^2 \frac{\partial^2 {\pi}}{\partial k^2} (k, t) = 0, 
\end{align} 
subject to the rescaled initial and boundary conditions
\begin{align}
  \label{eqn bc_call}
    {\pi}^c (k, 0) = \big( S_0 - K_\text{max} k \big)^+ , \quad {\pi}^c (\infty, t) = 0,
\end{align}
for call option prices, and
\begin{align}
  \label{eqn bc_put}
    {\pi}^p (k, 0) = \big( K_\text{max} k - S_0 \big)^+ , \quad
    {\pi}^p (0, t) = 0, 
\end{align}
for put option prices, as well as the inequality constraints \eqref{eqn inequality}. We note the conditions 
\begin{align}
\label{strikearb}
\frac{\partial \pi}{\partial T}(K,T) \propto
\frac{\partial {\pi}}{\partial t} (k, t) - rT_\text{max} k \frac{\partial {\pi}}{\partial k} (k, t)
 \geq 0,    
\end{align}
and
\begin{align}
\label{calendararb}
\frac{\partial^2 \pi}{\partial K^2}(K,T) \propto
 \frac{\partial^2 {\pi}}{\partial k^2} (k,t) \geq 0,
\end{align}
for the absence of strike and calendar arbitrage opportunities
respectively, see, e.g. \S2.2.2 of \cite{bergomi}. 
\begin{lemma}
  \label{jklsd1} 
 Conditions~\eqref{strikearb}
 and
 \eqref{calendararb}
 for the absence of strike and calendar arbitrage can be
 simultaneously enforced under the single inequality
\begin{align}
  \label{eqn arbitrage-free}
    f_\text{arb} (k, t) := \frac{\partial {\pi}}{\partial t} (k, t) - rT_\text{max} k \left( \frac{\partial {\pi}}{\partial k} (k, t) \right)^+  \geq 0.
\end{align}
\end{lemma}
\begin{Proof}
  We consider two cases.
\\
  \noindent
  $(i)$ 
  $\partial {\pi} / \partial k \geq 0$.
  In this case, 
  \eqref{eqn arbitrage-free} is clearly equivalent to
  \eqref{strikearb}.
  In addition, if \eqref{strikearb} holds
  then \eqref{eqn Dupire_pde_dimensionless}
  shows that \eqref{calendararb} is satisfied, as 
 $$
        {\eta} (k,t) k^2 \frac{\partial^2 {\pi}}{\partial k^2} (k, t) =
        \frac{\partial {\pi}}{\partial t} (k ,t)
        \geq rT_\text{max} k \frac{\partial {\pi}}{\partial k} (k, t) \geq 0. 
        $$
        \noindent
        $(ii)$ $\partial {\pi} / \partial k < 0$.
        In this case,
 \eqref{eqn arbitrage-free} becomes
 \begin{equation}
   \label{jfdgf} 
 \frac{\partial {\pi}}{\partial t} (k, t) \geq 0,
\end{equation} 
 which by
 \eqref{eqn Dupire_pde_dimensionless}
 is equivalent to \eqref{calendararb}.
 In addition, if \eqref{jfdgf} holds
 then \eqref{strikearb} is satisfied. 
 \end{Proof}
\subsection{Model and loss function} \label{sec:loss}

Market option prices are observed at discrete strike-maturity pairs, whereas it is favorable to have an option price and a local volatility that span continuous surfaces over the extended regimes of strike price and maturity. To obtain a continuous limit, we propose to model both the option price and the local volatility using neural networks. The associated neural network ansatz and the corresponding loss functions are detailed below.

\subsubsection{Neural network ansatz}

As option prices may vary dramatically against strike prices, we model the option price surface as an exponential function where the exponent is approximated using a neural network, whose model parameters $\boldsymbol{\theta}$ are learned from the market data.
We consider the neural ansatz 
\begin{align}
  \label{eqn neural_call}
    {\pi}^c_{\theta}(k, t) = S_0 \big(1 - e^{ -(1-k) \mathcal{N}_{c}(k,t; \boldsymbol{\theta}_{c})} \big),
\end{align}
and 
\begin{align}
   \label{eqn neural_put}
    {\pi}^p_{\theta}(k, t) = K_\text{max} e^{rT_\text{max} t} k \big(1- e^{ -k \mathcal{N}_{p}(k,t; \boldsymbol{\theta}_{p})}\big),
\end{align}
for call and put option prices, respectively.
At the practical infinity $k=1$ and at $k=0$, the boundary conditions for call \eqref{eqn bc_call} and for put \eqref{eqn bc_put} options are satisfied by construction. Here, a softplus function, $\text{softplus}(x) := \log(1 + e^x)$ is selected to be the activation function at the output layer of neural networks, so that the inequality constraints \eqref{eqn inequality} are always satisfied. 

Unlike for option prices, the variation of local volatility is relatively small, hence
 we model it using the neural network 
\begin{align}
  \label{eqn neural_vol}
        {\eta}_\theta (k, t) = \mathcal{N}_{\eta} (k, t; \boldsymbol{\theta}_{\eta}), \quad \mathcal{N}_{\eta}: \mathbb{R}_{+} \times \mathbb{R}_{+} \to \mathbb{R}_{+},
\end{align}
with a softplus activation function at the output layer,
so that ${\eta}_\theta (k, t)$ is non-negative by construction.
By modeling the squared local volatility using a neural network we also ensure
the smoothness of local volatility,
therefore mitigating numerical instabilities. 

Throughout the paper, we use the same neural network architecture for $\mathcal{N}_{c}$,
 $\mathcal{N}_p$ and $\mathcal{N}_{\eta}$. The network consists of three residual blocks \cite{He16}, where the residual connection is used around two sub-layers made of $64$ neurons and regularized with batch normalization \cite{Ioffe15} per layer. The model parameters of neural networks $\boldsymbol{\theta}_c$ (or $\boldsymbol{\theta}_p$) and $\boldsymbol{\theta}_\eta$ are determined by minimizing properly designed loss functions to be discussed in Section~\ref{sec: loss functions}. Since evaluating loss functions associated with the arbitrage-free conditions and with the Dupire equations requires computing successive derivatives, the $\tanh$ activation function is selected for hidden network layers.

\subsubsection{Loss functions} \label{sec: loss functions}

Given $N$ market data triplets $(\pi_i, K_i, T_i)$ with $i = 1, ..., N$,
we transform $K_i$ and $T_i$ using Eq.~\eqref{eqn change_of_variables},
leading to $({\pi}_i, k_i, t_i)$,
and consider the following loss functions.
\begin{enumerate}[i)]
\item
  The data loss arising from fitting market option prices using neural networks
  is defined as the squared error 
\begin{align}
\nonumber
      L_\text{fit} = \frac{1}{N} \sum_{i=1}^{N} w(\pi_i) \big( {\pi}_{\theta}(k_i, t_i) - {\pi}_i \big)^2 
\end{align}
 with the weight function
 \begin{align}
   \nonumber 
 w(\pi_i) : = 1 + \frac{N g(\pi_i)}{\sum_{j=1}^{N} g(\pi_j)},
\end{align}
 where
 \begin{align}
   \nonumber 
    g(x_i) := \frac{1}{10} \mathds{1}_{[0,0.1)}\left( x_i \right) +
  x_i \mathds{1}_{[0.1,10]}(x_i) + 10 \mathds{1}_{[10,\infty)}(x_i), 
\end{align}
 and 
 \begin{align}
   \label{eqn weight_x}
  x_i = \text{stop\_gradient} (\tilde{x}_i), \quad \mbox{with} \quad \tilde{x}_i = 
   \frac{1}{N \pi_i^2}\sum_{j=1}^{N} \pi_j^2, 
\end{align}
so that squared and relative squared errors are minimized. Here,
indicator functions
are used to mitigate numerical instabilities associated with $\pi_i = 0$,
 while leveraging contributions from large values of $\pi_i$.
 In addition, in \eqref{eqn weight_x} the operator $\text{stop\_gradient}(\cdot)$
 prevents the input variables of indicator functions from being
 taken into account when computing gradients,
 hence the weight function $w(\cdot)$ is treated as a scalar quantity during
 backpropagation.
 As the weight function
 $w$ is designed to balance losses evaluated at each collation point, it
 will be included in all loss functions to be defined in the following. 
 \item 
 As the initial conditions in Eqs. (\ref{eqn bc_call}) and (\ref{eqn bc_put}) are piecewise linear and not differentiable, 
  instead of imposing initial conditions as hard constraints into the neural ansatz
 we choose to minimize the loss function
\begin{align}
\nonumber
      L_\text{ini} = \frac{1}{M_1} \sum_{j=1}^{M_1} w\left(\pi(k_j, 0)\right) \big( {\pi}_\theta (k_j, 0) - {\pi}(k_j, 0) \big)^2,
\end{align}
 where the reference function ${\pi}(k_j, 0)$ is defined in Eq.~\eqref{eqn bc_call} for call options, and in Eq.~\eqref{eqn bc_put} for put options. To evaluate $L_\text{ini}$, $M_1$ synthetic collocation points are uniformly sampled from the interval $k_j \in [0,1]$ at each iteration during the minimization.
 \item 
   To ensure that the parameterized option price surface is arbitrage-free, on $M_2$ synthetic collocation points $(k_j, t_j)$ uniformly sampled from $\Omega := [0,1] \times [0,1]$, we substitute $\pi_\theta(k_j,t_j)$ into the function $f_\text{arb}(k, t)$ defined in Eq.~\eqref{eqn arbitrage-free}, and penalize the locations where $f_\text{arb} (k_j, t_j)$ is negative,
    which results into the loss function 
\begin{align}
\nonumber
      L_\text{arb} = \frac{1}{M_2} \sum_{j=1}^{M_2} w\left( \frac{\partial \pi_\theta}{\partial t} (k_j, t_j) \right) ( ( -f_\text{arb} (k_j, t_j) )^+ )^2 
\end{align}
associated with the arbitrage-free conditions.
\item 
Under the assumption that the price process of underlying assets of the option matches the local volatility model (\ref{eqn BS}), ${\pi}_\theta (k, t)$ must be a solution to the rescaled Dupire's equation (\ref{eqn Dupire_pde_dimensionless}) up to some error terms. The self-consistency between the parameterized option price and the calibrated local volatility is therefore established by minimizing the residue $f_\text{dup}(k, t)$ defined in Eq.~\eqref{eqn Dupire_pde_dimensionless}, leading to
\begin{align}
\nonumber
    L_\text{dup} = \frac{1}{M_2} \sum_{j=1}^{M_2} w
  \left( \frac{\partial \pi_\theta}{\partial t} (k_j, t_j)
  \right) ( f_\text{dup}(k_j, t_j) )^2,
\end{align}
where, similar to $L_\text{arb}$, the loss function arising from the Dupire equation are evaluated from the ensemble of uniformly sampled $M_2$ collocation points.
\end{enumerate}
 In addition, the initial condition provides reference option prices on collocation points sampled on the line $T=0$, equivalent to the quoted market option prices.
A successive re-sampling then covers the entire support domain $\Omega$, ensuring that both the arbitrage-free condition and the Dupire equation are everywhere satisfied. This not only avoids the need to design an adaptive mesh for Dupire's equation as in \cite{Achdou05}, but also mitigates overfitting that can be due to the uneven distribution of scarce data.
 Computing loss functions $L_\text{arb}$ and $L_\text{dup}$ involves evaluating differential operators, which is conveniently achieved by automatic differentiation \cite{Baydin17}.

Finally, the joint minimization of 
\begin{align}
\label{eqn loss_pi}
    L_{\pi} = L_\text{fit} + \lambda_\text{ini} L_\text{ini} + \lambda_\text{arb} L_\text{arb} + \lambda_\text{dup} L_\text{dup}, 
\end{align}
where $\lambda_\text{ini}, \lambda_\text{arb}, \lambda_\text{dup} \in \mathbb{R}_+$, with respect to $\boldsymbol{\theta}_c$ (or $\boldsymbol{\theta}_p$) brings correction to the observed market data,
 by filtering out noises that may lead to arbitrage opportunities or violate the price dynamics underpinned by the scaled Dupire's equation. 
 Without loss of generality, we take $\lambda_\text{ini} = \lambda_\text{arb} = 1$, reducing the number of hyper parameters to one: $\lambda_\text{dup} = \lambda \in \mathbb{R}_{+}$.
 
 On the other hand, the minimization of $L_\text{dup}$ with respect to $\boldsymbol{\theta}_\eta$ ensure 
  consistency between the local volatility function estimate and
 the parameterized option price.
  In this sense, the proposed method is self-consistent. 
To assess the robustness of the proposed self-consistent method, various values of $\lambda$ are considered in Section~\ref{sec: case_study} as an ablation study.

\subsection{Algorithm} \label{sec:algo}

Based on the deep self-consistent learning method discussed above,
Algorithm~\ref{algo} outlines a computational scheme for the
 calibration of local volatility. 
 At variance with usual approaches where the derivatives of the option price are first approximated to calibrate the local volatility
 using Dupire's formula \eqref{eqn Dupire_formula},
 this algorithm approximates both the option price and the local volatility using neural networks.

 \begin{algorithm}[H]
    \caption{Deep self-consistent learning for option prices and local volatility.}
    \label{algo}
    \textbf{Input:} Scaled market call (or put) option prices and transformed strike-maturity pairs $({\pi}_i, k_i, t_i)$ with $i = 1, ..., N$. \\
    Guess initial parameters $\boldsymbol{\theta}_c$ (or $\boldsymbol{\theta}_c$) and $\boldsymbol{\theta}_{\eta}$.
    \begin{algorithmic}[1]
    \While{not converged}{ 
    \State Uniformly draw $M_1$ points $k \in [0,1]$ and $M_2$ scaled strike-maturity pairs $(k, t) \in \Omega$. 
    \State Compute $4$ loss functions: $L_\text{fit}$, $L_\text{ini}$, $L_\text{arb}$, $L_\text{dup}$, and their weighted sum $L_\pi$.
    \State Optimize $\boldsymbol{\theta}_c$ (or $\boldsymbol{\theta}_p$) using gradients $\partial_{\boldsymbol{\theta}_{\pi}} L_\pi$. 
    \State Optimize $\boldsymbol{\theta}_\eta$ using gradients $\partial_{\boldsymbol{\theta}_\eta} L_\text{dup}$.
    \EndWhile}
    \end{algorithmic}
    \hspace*{\algorithmicindent} \textbf{Output}: Optimized parameters $\boldsymbol{\theta}_c$ (or $\boldsymbol{\theta}_p$) and $\boldsymbol{\theta}_{\eta}$.\\
    Hyper-parameters used in this paper: $M_1 = 128$, $M_2 = 128^2$, $\lambda_\text{ini} = \lambda_\text{arb} = 1$ and $\lambda_\text{dup} = \lambda$.
  \end{algorithm}

 \noindent 
 The training of neural networks is a non-convex optimization problem which
 does not guarantee convergence towards the global minimum, hence regularization is required
 and provided via the residue of the scaled Dupire equation
 in the term $L_\text{dup}$.
 The joint minimization of $L_\text{fit}$ arising from the deviation of the parameterized option price to the market data, $L_\text{ini}$ due to the initial conditions, $L_\text{arb}$ linked with the arbitrage-free conditions, and $L_\text{dup}$
 seeks a self-consistent pair of approximations for the option price and for the underlying local volatility, and
  to exclude local minima that violate these constraints.

  Moreover, in practice, strike-maturity pairs with quoted option prices are usually unevenly distributed \cite{Ackerer20,Chataigner21},
  hence a direct minimization of loss functions evaluated on the quoted strike-maturity pairs
  may yield large calibration errors of local volatility at locations where the measurement is scarce.
  On the other hand,
  being continuous functions, neural networks
  allow for a mesh-free discretization of Dupire's equation by uniformly sampling collocation points from $\Omega$ at each iteration,
  therefore improving calibration from scarce and unevenly distributed data.
  Lastly, since both option price and squared local volatility are approximated using neural networks, the smoothness of the option price and local volatility surface are guaranteed, 
  while positiveness is ensured by a proper choice of output activation functions.
  As a result, the
  numerical instabilities associated with the calibration
  of local volatility using
  the Dupire formula \eqref{eqn Dupire_formula} are less likely to occur. 

\section{Case studies} \label{sec: case_study}

\begin{figure}[H]
  \centering
  \noindent\makebox[\textwidth]{
  \includegraphics[width=1.22\textwidth]{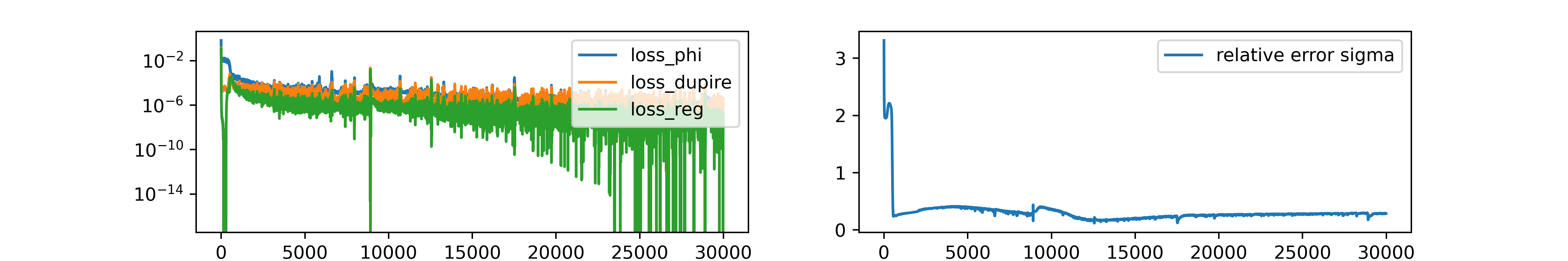}}
\vspace{-0.5cm} 
\caption{\small Left: violations of the non-arbitrage condition; and right: relative local volatility error.}
\label{fig:l-arb1} 
\end{figure}

In this section, we test our self-consistent method, first on synthetic option prices in Section~\ref{sec:synthetic_data}, and then on real market data in Section~\ref{sec:real_data}.
 We start the training with an initial learning rate $10^{-3}$, which is divided by a decaying factor $1.1$ every $2,000$ iterations. On a workstation equipped with an Nvidia RTX 3090 GPU card, each iteration takes less than $0.01$ second. The total number of iterations is capped at $30,000$ such that the overall computational time is compatible with that in \cite{Chataigner20,Chataigner21}, enabling a fair comparison. 
 Figure~\ref{fig:l-arb1} presents a typical graph of loss functions
 $L_{\rm ini}$,
 $L_{\rm dup}$
 and 
 $L_{\rm arb}$
 observed in the following experiments  
 as the number of iterations increases.
  
\vspace{-0.3cm} 
 
\vspace{-0.3cm} 
 
\begin{figure}[H]
  \centering
  \noindent\makebox[\textwidth]{
    \includegraphics[width=0.4\textwidth]{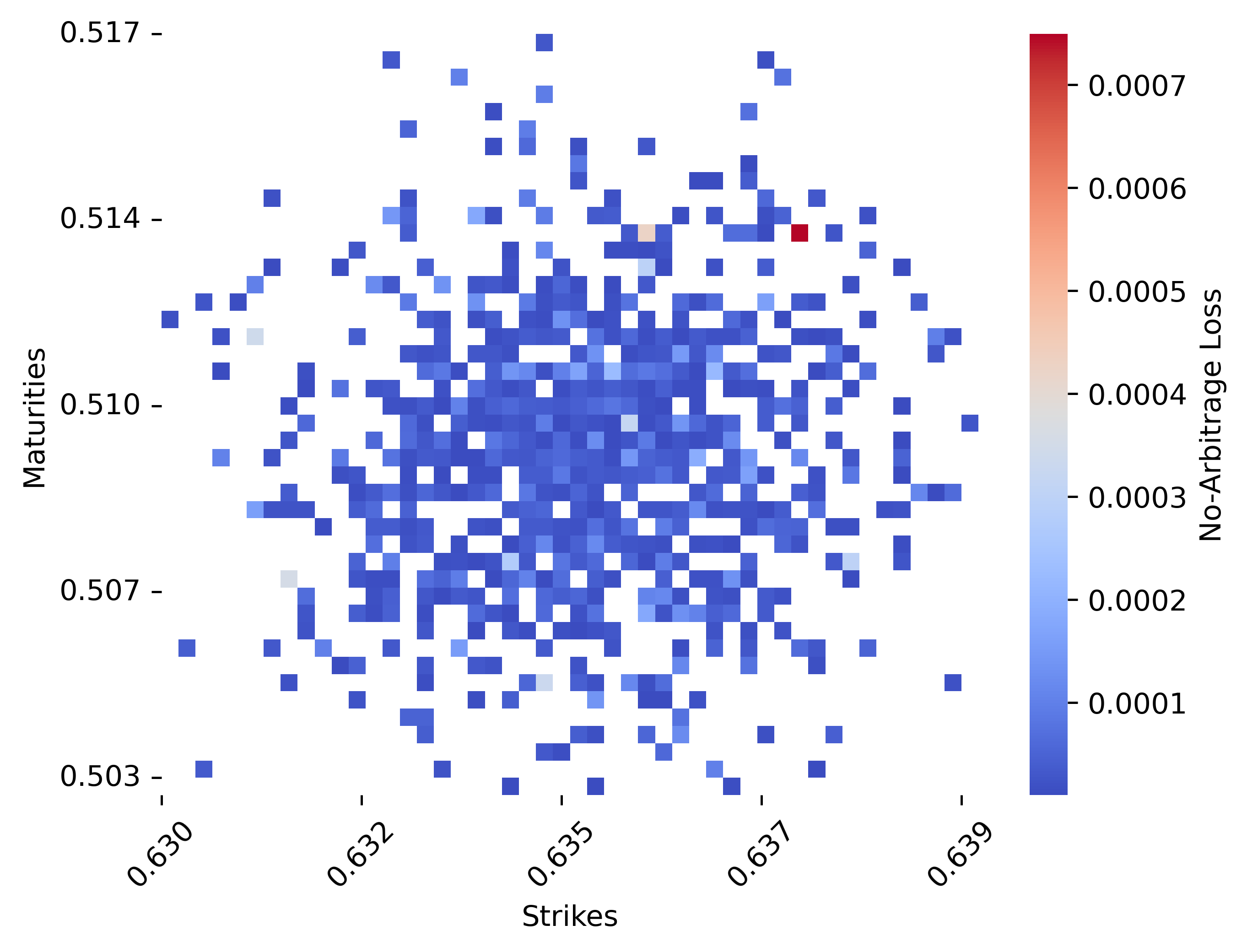}
    \hskip1.cm
    \includegraphics[width=0.4\textwidth]{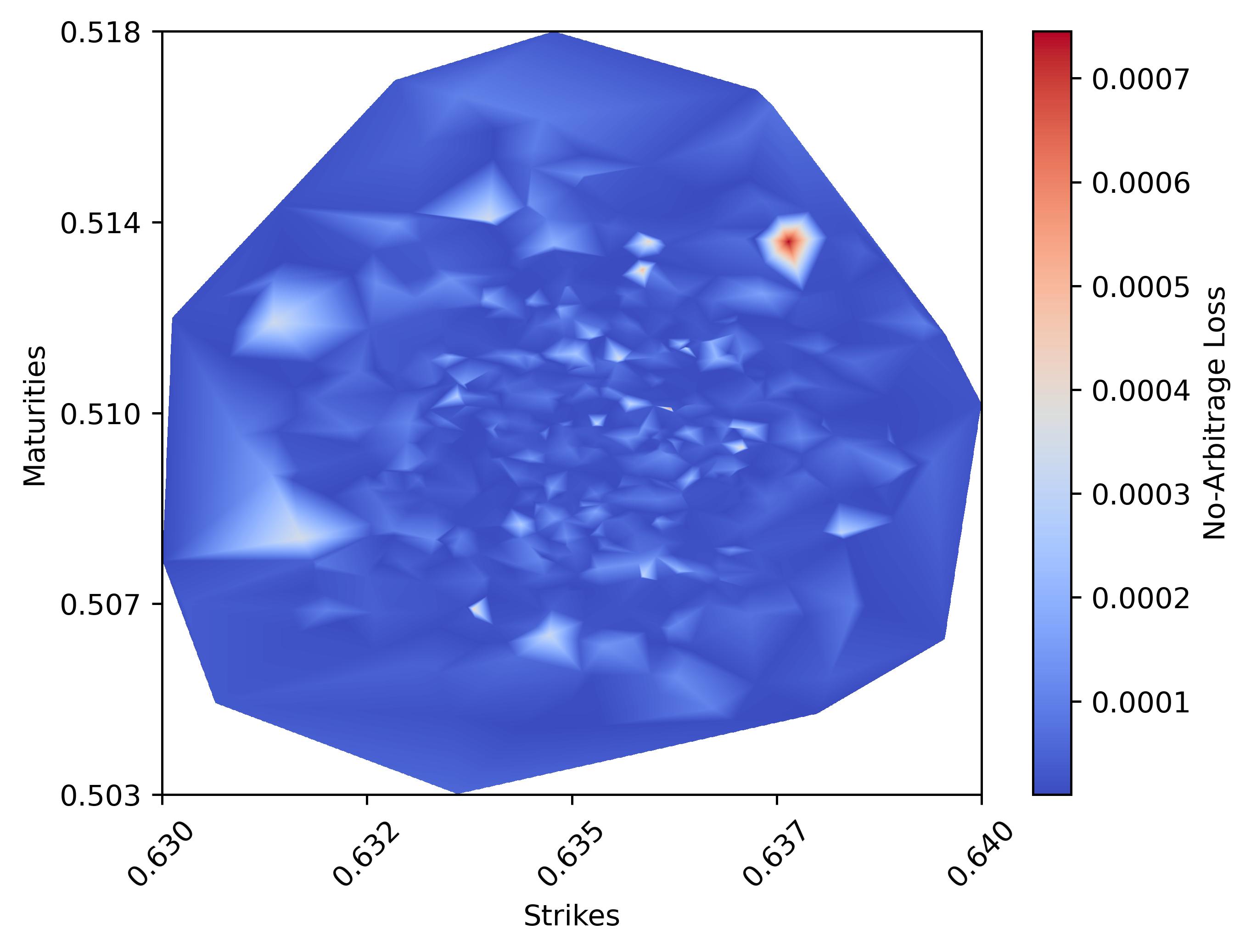}
}
\vspace{-0.6cm} 
\caption{\small Left: heatmap of non-arbitrage condition violations; and right: linear interpolation.}
\label{fig:l-arb2} 
\end{figure}

   \noindent
   Figure~\ref{fig:l-arb2} presents a heatmap of the
   function $f_\text{arb}$, showing the 
   no-arbitrage violations at various collocation points. 
  
\vspace{-0.cm} 
 
\vspace{-0.7cm} 
 
\subsection{Synthetic data for European call options}
 \label{sec:synthetic_data}
Synthetic option prices are generated as Monte Carlo estimates of Eq.~\eqref{eqn expectation_option} for European call options at given strike-maturity pairs $(K_i, T_i)$. Here, the asset price paths $S_t$ are obtained by simulating the local volatility model \eqref{eqn BS}
with $r = 0.04$ and
\begin{align}
  \label{eqn sigma_exact}
  \sigma(x,t) = 0.3 + y e^{-y} \quad \mbox{with} \quad y = (t + 0.1) \sqrt{x+0.1},
 \end{align}
making it possible to assess the validity of the calibrated local volatility
 by comparison with the closed-form expression of $\sigma(x,t)$.
 The price trajectories are obtained by simulating Eq.~\eqref{eqn BS} for $10^6$ times from a single initial condition $S_0 = 1,000$ for the period $t = [0, 1.5]$ and with a time step $\mathrm{d} t = 0.01$.
 \subsubsection*{Large data set}
\begin{figure}[H]
  \centering
  \noindent\makebox[\textwidth]{
  \includegraphics[width=\textwidth]{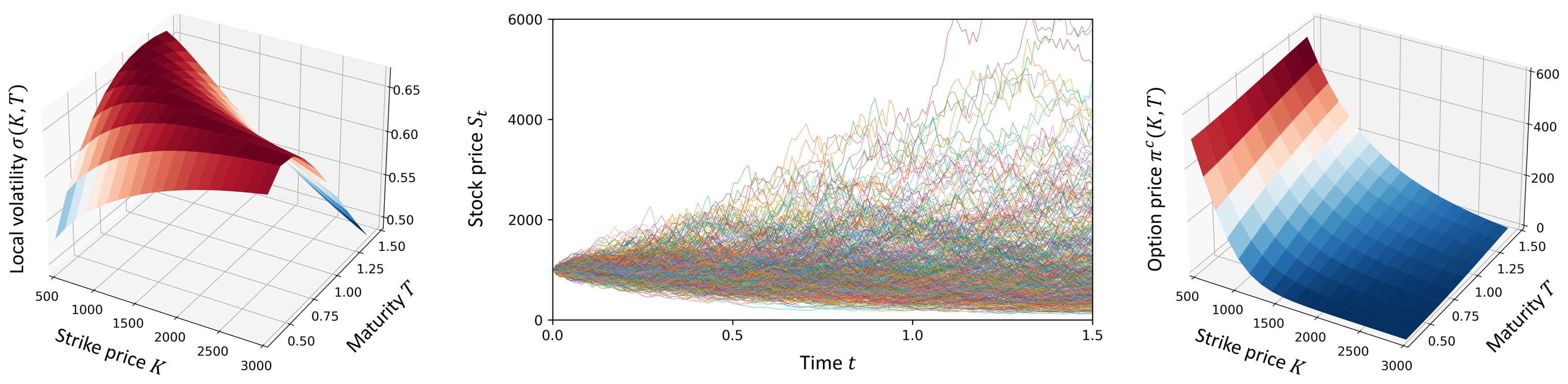}}
  \caption{\small Left: exact local volatility surface, cf. Eq.~\eqref{eqn sigma_exact}; middle: simulated price trajectories by solving the local volatility model. cf. Eq.~\eqref{eqn BS}; and right: generated call option price using Monte Carlo simulation of Eq.~\eqref{eqn expectation_option} with $10^6$ sampled price trajectories. The local volatility and the option price surfaces are visualized on an evenly spaced $10 \times 20$ grids and, for a better visualization, only the first $1024$ price trajectories are shown in the middle pane. }
  \label{fig:synthetic_data}
\end{figure}

 To begin with, we price European call options on a mesh grid consisting of evenly spaced $10$ points for the $[0.3, 1.5]$ $T$-period and $20$ points in the $[500, 3000]$ $K$-interval, respectively. That is, the dataset consists of $10 \times 20$ option prices quoted at the corresponding strike-maturity pairs, leading to the triplet $(\pi^c_i, K_i, T_i)$ with $i = 1, ..., 200$. The exact local volatility surface, the simulated price trajectories by numerically integrating the local volatility model, and the synthetic call option price surface are visualized in Figure~\ref{fig:synthetic_data}.
 
\vspace{-0.3cm} 
 
\begin{figure}[H]
  \centering
  \noindent\makebox[\textwidth]{
  \includegraphics[width=\textwidth]{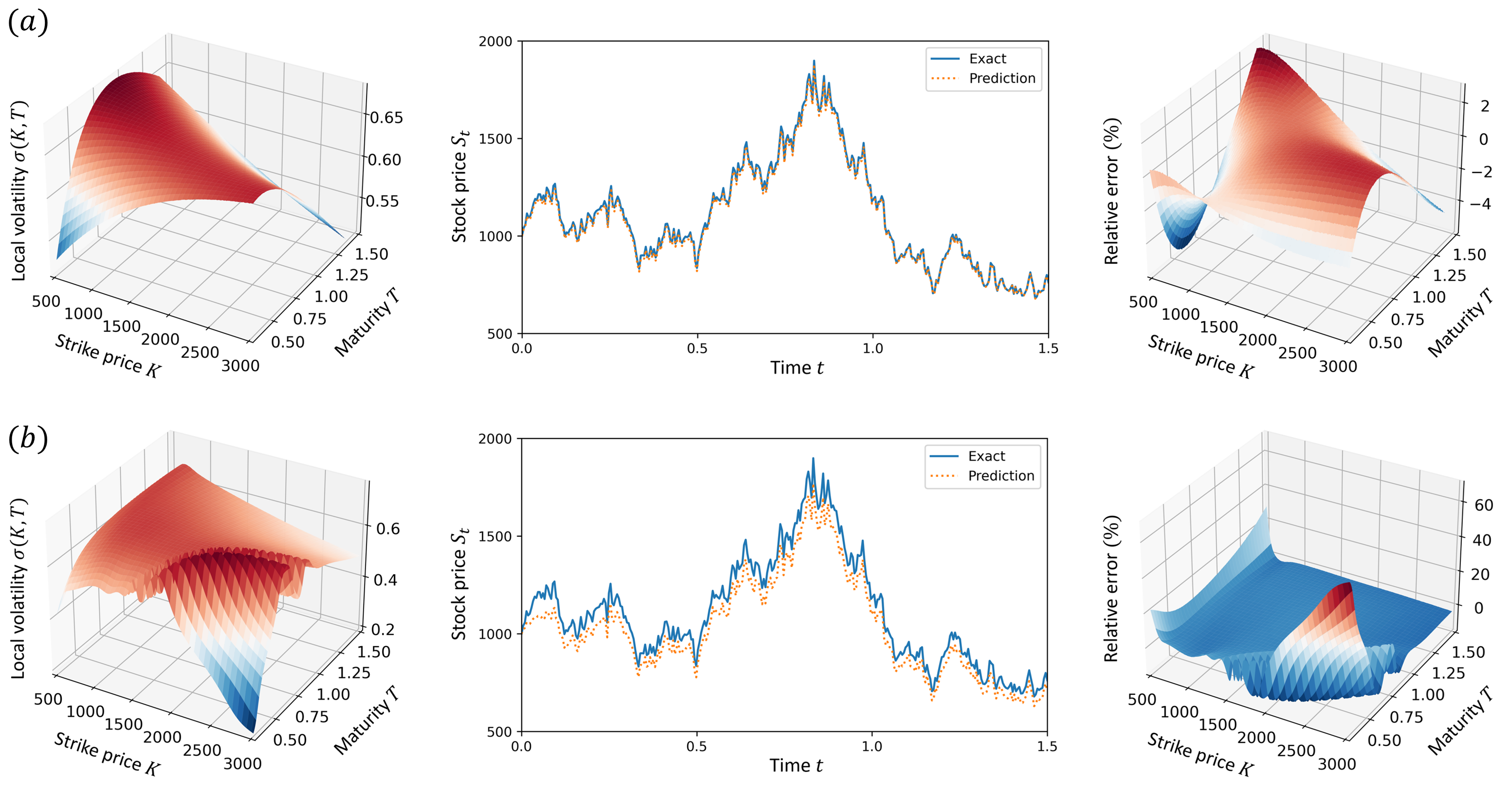}}
  \caption{\small Visualization of calibrated neural local volatility (left), simulated price trajectory with a pre-sampled Brownian motion $(B_t)$  (middle), and the relative error of the calibrated local volatility with respect to the exact expression \eqref{eqn sigma_exact} $(a)$ with ($\lambda_\text{dup} = 1$); and $(b)$ without ($\lambda_\text{dup} = 0$) the inclusion of $L_\text{dup}$ as regularizer. The models shown in are trained on a dataset consists of $10 \times 20$ option prices, while the visualization and relative errors are computed on a different test grid with linearly spaced $256 \times 256$ collocation points.}
  \label{fig:synthetic_res}
\end{figure}

\noindent
After $\boldsymbol{\theta}_{c}$ and $\boldsymbol{\theta}_{\eta}$ are determined,
the parameterized option price and calibrated local volatility are recovered
from Eqs. \eqref{eqn neural_call} and \eqref{eqn neural_vol}. 
Thereafter, we substitute the calibrated neural local volatility into Eq.~\eqref{eqn BS}, forming a neural local volatility model, whose solution, in turn, gives the synthetic stock prices, from which the option price can then be recovered using Monte Carlo estimates of Eq.~\eqref{eqn expectation_option}, completing a control loop for model assessment. The calibrated local volatility, the simulated price trajectory by numerically integrating the neural local volatility model with a pre-sampled sequence of the Brownian motion $(B_t)$, and the relative errors of calibrated local volatility with respect to the exact expression \eqref{eqn sigma_exact} are shown in Figure~\ref{fig:synthetic_res}.

\vspace{-0.3cm}

\noindent
For $\lambda_\text{dup} = 0$, the parameterized option price is not regularized by a priori physics information encoded in Dupire's equation, reducing the self-consistent calibration of local volatility to a one-way approach. In both cases $(a)$-$(b)$ the parameterized option prices are regularized by the arbitrage-free conditions as in previous works \cite{Ackerer20,Chataigner20,Chataigner21}.
 On the other hand, exploiting self-consistency by the inclusion of $L_\text{dup}$ as a regularizer improves significantly the calibrated local volatility surface.
\subsubsection*{Small data set}
\begin{figure}[H]
  \centering
  \noindent\makebox[\textwidth]{
  \includegraphics[width=\textwidth]{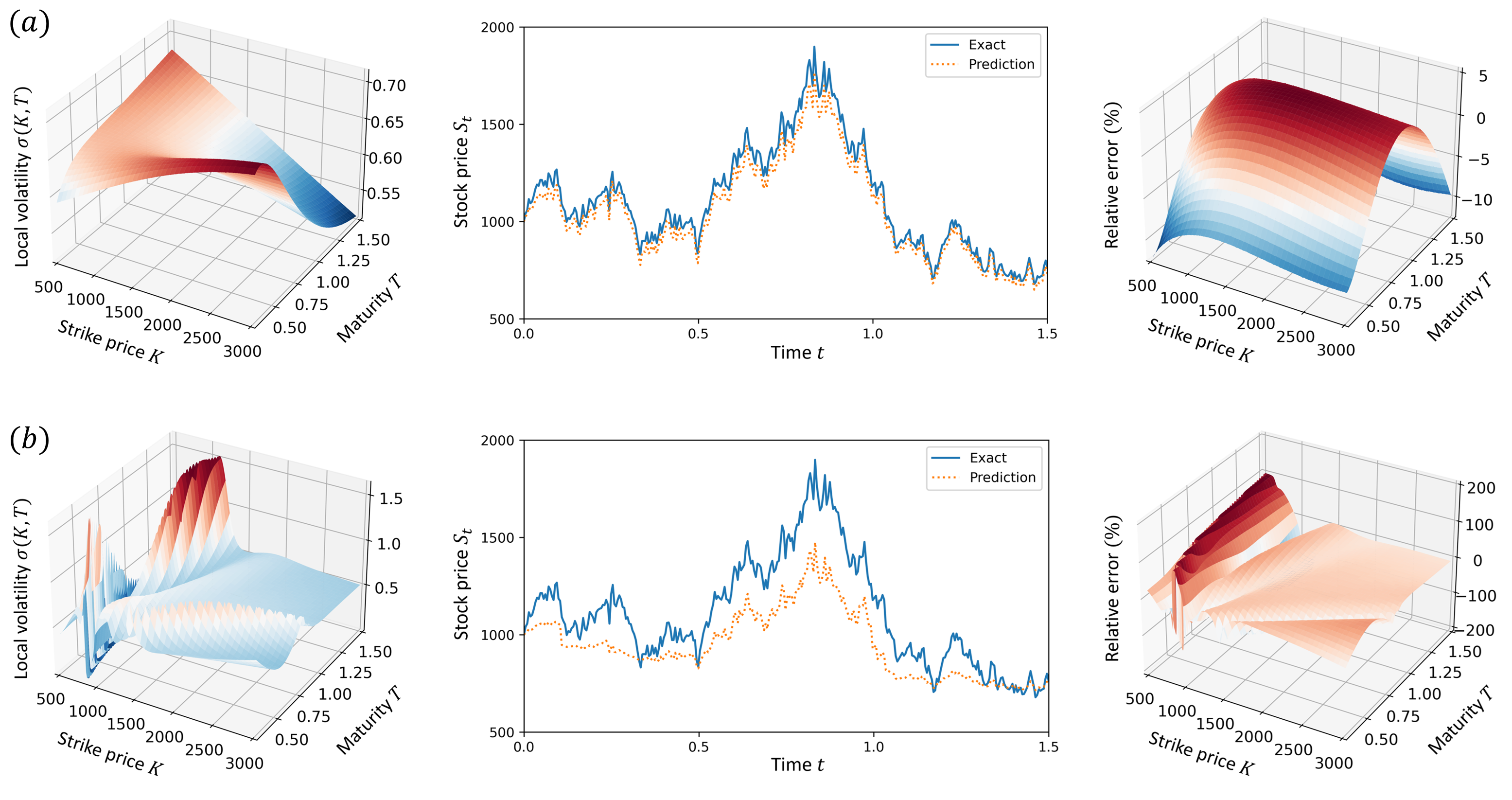}}
  \caption{\small Same as in Figure~\ref{fig:synthetic_res}, but with the models trained on a scarce dataset consisting of $3 \times 6$ option prices quoted on a linearly spaced grid $(K,T) \in [500, 3000] \times [0.3, 1.5]$. }
  \label{fig:synthetic_res_scarce}
\end{figure}

To explore the small data limit of the proposed method, we present in Figure~\ref{fig:synthetic_res_scarce} the calibrated local volatility surface from a scarce dataset consisting of $3 \times 6$ option prices quoted on a linearly spaced grid $(K,T) \in [500, 3000] \times [0.3, 1.5]$.

\vspace{-0.3cm}

\noindent
It is observed that the use of self-consistency effectively compensates for the lack of data, validating the proposed method in the small data limit. Indeed, neural networks are essentially nonlinear functions. Given discrete pairs of inputs and outputs, such a mapping function is not uniquely determined. To ensure that the obtained neural networks can be generalized, one needs to go through the entire domain of definition of neural networks, calling for big data. Alternatively, incorporating a priori physics information, completely or partially known, into the learning scheme can regularize neural networks at places where there is no data, allowing one to benefit from recent advances in deep learning without big data.
\subsubsection*{Sensitivity with respect to $\lambda_\text{dup}$} 
 \noindent
 To investigate the sensitivity of the proposed method with respect to $\lambda_\text{dup}$, we consider a sequence of values in the range $\lambda_\text{dup} = [0, 4]$. We quantify the accuracy of our self-consistent method by the root mean squared error (RMSE) for the parameterized option price and the
 RMSE for the calibrated local volatility,
  which is less relevant
   due to the non-uniqueness of the solution to such calibration problems.
 \label{skfd1}
   Here, independent of the training dataset, RMSEs are computed on a different testing grid consisting of a linearly spaced $256 \times 256$ collocation points in $[K, T] \in [500, 3000] \times [0.3, 1.5]$. We perform three independent runs for each case and the averaged results are summarized in Table~\ref{table:relative_error}. 

\begin{table}[H]
\centering
\renewcommand{\arraystretch}{1.75}
\begin{tabular}{|c|c|c|c|c|c|c|}
\hline
\multirow{2}{*}{Dataset}      & \multirow{2}{*}{RMSEs} & \multicolumn{5}{c|}{$\lambda_\text{dup}$}           \\ \cline{3-7} 
                                   &                         & $0.0$    & $0.5$     & $1.0$      & $1.5$      & $2.0$      \\ \hline

\multirow{3}{*}{$3 \times 6$}      & Option price            & $10.53$  & $2.27$    & $1.95$     & $1.62$     & $1.76$     \\ \cline{2-7} 
                                   & Local volatility        & $0.25$   & $0.03$    & $0.03$     & $0.02$     & $0.03$     \\ \cline{2-7} 
                                   & Reprice error           & $64.54$  & $1.84$    & $1.88$     & $1.81$     & $1.71$     \\ \hline

\multirow{3}{*}{$10 \times 20$}    & Option price            & $1.07$   & $1.76$    & $0.94$     & $1.10$     & $1.22$     \\ \cline{2-7} 
                                   & Local volatility        & $0.04$   & $0.01$    & $0.01$     & $0.01$     & $0.02$     \\ \cline{2-7} 
                                   & Reprice error           & $5.99$   & $1.23$    & $1.16$     & $1.08$     & $1.34$     \\ \hline
\end{tabular}
\caption{Test of proposed self-consistent method on synthetic dataset.}
\label{table:relative_error}
\end{table}

\noindent
 It is observed that, when increasing $\lambda_\text{dup}$ from $0$, the RMSEs for the calibrated volatility and the repriced option price first decrease then increase, indicative the existence of an optimal $\lambda_\text{dup}$. The value of optimal $\lambda_\text{dup}$ seems to be dependent on the size of dataset and its determination is out of the scope of this paper, hence left for future work. The decreased RMSEs for any reasonably chosen $\lambda_\text{dup} \neq 0$ support the inclusion of $L_\text{dup}$ for regularizing the parameterized option price. With randomly initialized $\boldsymbol{\theta}_\eta$, the corresponding Dupire equation forms essentially a wrong a priori for the measured data. During the training, a joint minimization of $L_\pi$ with respect to $\boldsymbol{\theta}_c$ and of $L_\text{dup}$ with respect to $\boldsymbol{\theta}_\eta$ yields a self-consistent pair of parameterized option price that provides an optimized description for the measured data, initial and boundary conditions, arbitrage-free conditions, and the underlying Dupire's equation; as well as the calibrated local volatility that matches, upon a physics-informed correction, the price dynamics of the observed data. Assigning a too small weight leads to unregularized parameterization, whereas a large $\lambda_\text{dup}$ slows down the minimization, resulting in an increased calibration error with a fixed number of iterations. 

\subsection{Application to market data} \label{sec:real_data}

In the case of market option prices, since the exact local volatility is not available, the assessment is performed by computing the reprice error. More specifically, we replace $\sigma(x,t)$ in Eq.~\eqref{eqn BS} by the calibrated one 
\begin{align}
  \label{eqn volatility}
    \sigma_\theta(x,t) = \sqrt{2{\eta}_\theta (x,t) / T_\text{max}}
\end{align}
and generate synthetic asset price paths via numerically integrating the neural local volatility model. The option is then repriced at given strike-maturity pairs using Monte Carlo estimates of Eq.~\eqref{eqn expectation_option} and,
 eventually, compared with the market call (or put) option prices.

 \subsubsection{European call options on the DAX index }

\begin{figure}[H]
\centering
  \noindent\makebox[\textwidth]{
  \includegraphics[width=\textwidth]{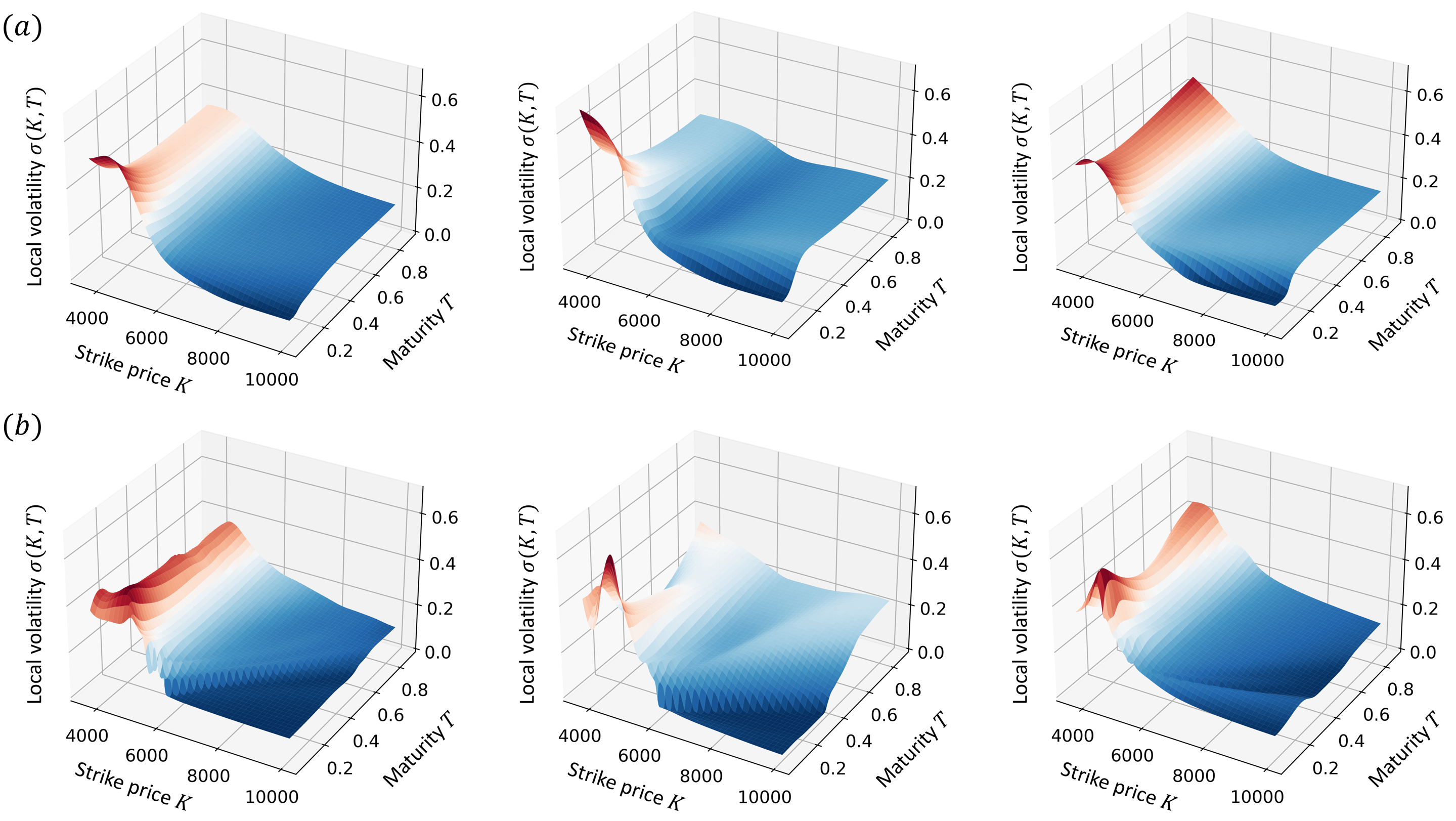}}
  \caption{\small Visualization for the neural local volatility surface calibrated $(a)$ with ($\lambda_\text{dup} = 1.0$); and $(b)$ without ($\lambda_\text{dup} = 0.0$) the inclusion of the residue of Dupire's equation as a regularizer. The obtained results are trained with DAX call option prices listed on $7$-th (left), $8$-th (middle), and $9$-th (right) August 2001. }
  \label{fig:DAX_res}
\end{figure}

 To perform a comparison with the literature \cite{Crepey02,Chataigner20}, we take the daily dataset of DAX index European call options listed on the $7$-th, $8$-th, and $9$-th, August 2001. The spot prices for the underlying assets are $S_0 = 5752.51, 5614.51, 5512.28$, respectively, with $217$, $217$, and $218$ call options quoted at differential strike-maturity pairs
 and $r = 0.04$. The calibrated local volatility surfaces with and without the inclusion of $L_\text{dup}$ as a regularizer are shown in Figure~\ref{fig:DAX_res}.
 
\vspace{-0.3cm}

\noindent
 It is observed that, comparing with the case $\lambda_\text{dup} = 0$, the inclusion of $L_\text{dup}$ achieves more stability of the local volatility surfaces over successive days. However, to determine whether the variations that we observe in Figure~\ref{fig:DAX_res} $(b)$ are linked with overfitting, we resort to the quantitative analysis below. 

 As in \cite{Chataigner20}, we assess the calibration of local volatility by computing the reprice RMSEs and the results are summarized in Table~\ref{table:repricing_rmse}.

\begin{table}[H]
\centering
\begin{threeparttable}
\renewcommand{\arraystretch}{1.75}
\begin{tabular}{|l|c|c|c|c|}
\hline
\multirow{2}{*}{\makecell{Reprice RMSE}}
& \multicolumn{1}{c|}{\makecell{Ours}}    & \multirow{2}{*}{\makecell{Tikhonov\\\cite{Crepey02}}} & \multicolumn{1}{c|}{\makecell{Ours}} & \multirow{2}{*}{\makecell{Price-based \\ neural network \\\cite{Chataigner20}}}
\\
\cline{2-2}  \cline{4-4}
& $\lambda_\text{dup} = 1$            &  & $\lambda_\text{dup} = 0$     &
\\ \hline
7 August 2001    & 2.85           & 5.42    & 9.44          & 10.18                  \\ \hline
8 August 2001     & 3.27           & 5.55   & 6.16           & 7.44                   \\ \hline
9 August 2001       & 2.67           & 4.60  & 8.93            & 8.18                   \\ \hline
\end{tabular}
\end{threeparttable}
\caption{Comparison of reprice RMSEs 
  on DAX call options.}
\label{table:repricing_rmse}
\end{table}

\vspace{-0.3cm} 

\noindent
 Our numerical results are averaged over three independent runs. The unregularized calibration, i.e. $\lambda_\text{dup} = 0$, leads to relatively large reprice RMSEs, evidencing that the observed variations of local volatility surface over successive days in Figure~\ref{fig:DAX_res} are indeed due to overfitting. On the other hand, compared with previous works using Tikhonov regularization \cite{Crepey02} and with regularizer that impose positiveness and boundedness of the local volatility \cite{Chataigner20}, the inclusion of $L_\text{dup}$ as a regularizer reduces significantly the reprice RMSEs, leading to the best results. This, therefore, asserts quantitatively the effectiveness of the proposed self-consistent approach.

\subsubsection{Put options on the SPX index}

\begin{figure}[H] 
  \centering
  \noindent\makebox[\textwidth]{
  \includegraphics[width=\textwidth]{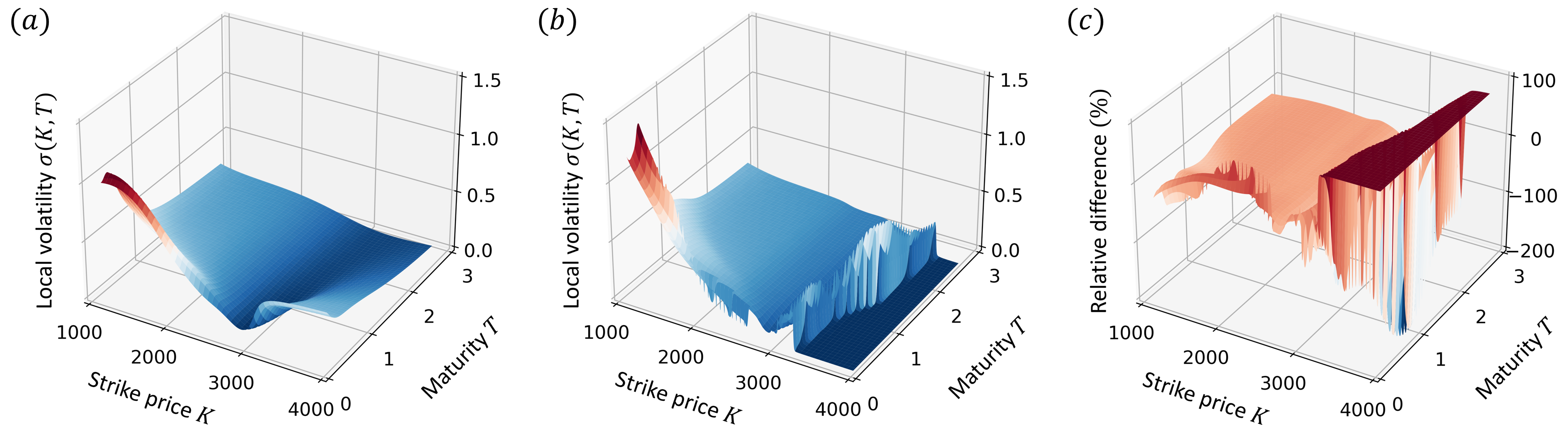}}
  \caption{\small Visualization for the neural local volatility surface calibrated $(a)$ with ($\lambda_\text{dup} = 1$); $(b)$ without ($\lambda_\text{dup} = 0$) the inclusion of the residue of Dupire's equation as a regularizer; and the relative difference of the local volatility surface obtained in $(b)$ computed with respect to that in $(a)$. The obtained results are trained with SPX call option prices listed on 18th May 2019. For better visualization, the relative difference in $(c)$ is truncated at $100\%$.}
  \label{fig:SPX_res}
\end{figure}

In our second application to market data, we use the same dataset, i.e. SPX European put options listed on 18th May 2019, as in \cite{Chataigner21}. The put options are quoted on maturities $T \in [0.055, 2.5]$ and with various strike prices $K \in [1150, 4000]$. Spot price of the underlying is $S_0 = 2859.53$ and $r = 0.023$. Following \cite{Chataigner21}, the dataset is split into a training dataset and a testing dataset consisting of $1720$ and $1725$ market put option prices, enabling a comparison. The calibrated local volatility surfaces for the cases $\lambda_\text{dup} = 0$ and $\lambda_\text{dup} = 1$, as well as their relative difference are shown in Figure~\ref{fig:SPX_res}. It is observed that a numerical singularity developed around $K=3000$ for the case $\lambda_\text{dup} = 0$ is suppressed with the inclusion of $L_\text{dup}$ as a regularizer.

\noindent
Next, on the testing dataset, we evaluate the interpolation and reprice RMSEs of the option prices with and without the inclusion of $L_\text{dup}$ as a regularizer.
 We then compare our results with published benchmarks obtained using
 the surface stochastic volatility inspired (SSVI) model,
 the Gaussian process (GP),
 as well as the implied volatility (IV) based
 and price based neural network methods,
 see \cite{Chataigner21} for details.
 In Table~\ref{table:repricing_rmse_spx}, the RMSEs are evaluated on the testing grid and our results shown are averaged over three independent runs.

\begin{table}[H]
\centering
\begin{threeparttable}
\renewcommand{\arraystretch}{1.75}
\begin{tabular}{|l|c|c|c|c|c|c|c|c|}
\hline
\multirow{2}{*}{\makecell{RMSEs}} & \multicolumn{2}{c|}{\makecell{Ours}} & \multirow{2}{*}{SSVI} & \multirow{2}{*}{GP}  & \multicolumn{2}{c|}{\makecell{Neural networks}} \\  \cline{2-3}\cline{6-7}
                 & $\lambda_\text{dup} = 0$    & $\lambda_\text{dup} = 1$    &             &            & IV  & Price  \\ \hline
Interpolation    & 0.97             & 1.92             & 2.89        & 0.26       & 2.97      & 10.35        \\ \hline
Reprice          & 5.24             & 3.26             & 22.83       & 74.02      & 4.99      & 11.76        \\ \hline
\end{tabular}
\end{threeparttable}
\caption{Comparison of interpolation and reprice RMSEs with \cite{Chataigner21} on SPX put options.}
\label{table:repricing_rmse_spx}
\end{table}

\vspace{-0.3cm}
 
\noindent 
With $\lambda_\text{dup} = 1$, our proposed method achieves the lowest reprice RMSEs on the testing dataset, confirming the significance of the proposed self-consistent approach. We note that the Gaussian process, which achieves the lowest interpolation error among all methods, results in the highest reprice RMSEs. This can indicate that the market data of the put option prices contain a significant amount of noise, which may be linked to the singular behavior observed in Figure~\ref{fig:SPX_res}$(b)$, leading eventually to the high reprice RMSEs in Table~\ref{table:repricing_rmse_spx}. Consequently, compared with the case $\lambda_\text{dup} = 0$, a simultaneous increase in interpolation RMSE and decrease in reprice RMSEs in the case $\lambda_\text{dup} = 1$ may imply a physics-informed correction to the market data via self-consistency.

\section{Discussion and conclusions} \label{sec:limitations}

In this work, we introduce a deep learning method that yields the parameterized option price and the calibrated local volatility in a self-consistent manner. More specifically, we approximate both the option price and the local volatility using deep neural networks. Self-consistency is established through Dupire's equation in the sense that the parameterized option price from the market data is required to be a solution to the underlying Dupire's equation with the calibrated local volatility. Consequently, by exploiting self-consistency, one not only calibrates the local volatility surface from the market option prices, but also filters out noises in the data that violate Dupire's equation with the calibrated local volatility, going beyond classical inverse problems with regularization.

The proposed method has been tested on both synthetic and market option prices. In all cases, the proposed self-consistent method results in a smooth surface for the calibrated local volatility, with the reprice RMSEs are lower than that obtained either by the canonical methods or the deep learning approaches \cite{Crepey02,Achdou05,Chataigner20,Chataigner21}. Moreover, the reprice RMSEs are relatively insensitive to the regularization parameter $\lambda_\text{dup} > 0$, showing the robustness of our algorithm. Being continuous functions, the neural networks provide full surfaces for the parameterized option price and for the calibrated local volatility, at variance with discrete nodes using the canonical methods. However, incorporating the residue of Dupire's equation as a regularizer requires one to solve a two-dimensional partial differential equation at each iteration, leading to increased computation time. This drawback, however, can be mitigated by distributing the training task on multiple GPUs. 

Although option prices may vary dramatically due to the price dynamics of the underlying assets, it is observed from Figure~\ref{fig:DAX_res} that the variation of the local volatility surface over successive days remains small.
Therefore, instead of starting from random initial parameters, initiating the training from converged solutions of the previous day can lead to a substantial reduction in computation time, see \cite{WangZhe21} for a case study.

\section*{Acknowledgment}
Zhe Wang would like to thank supports from Energy Research Institute@NTU, Nanyang Technological University, 
 where part of the work was performed.

\footnotesize

\def\cprime{$'$} \def\polhk#1{\setbox0=\hbox{#1}{\ooalign{\hidewidth
  \lower1.5ex\hbox{`}\hidewidth\crcr\unhbox0}}}
  \def\polhk#1{\setbox0=\hbox{#1}{\ooalign{\hidewidth
  \lower1.5ex\hbox{`}\hidewidth\crcr\unhbox0}}} \def\cprime{$'$}

\end{document}